*Quantitative Imaging of Protein-Protein Interactions by Multiphoton Fluorescence Lifetime Imaging Microscopy using a Streak camera*

*R. V. Krishnan,\*  A. Masuda, V. E. Centonze and B. Herman*

Department of Cellular and Structural Biology, University of Texas Health Science Center, 7703 Floyd Curl Drive, San Antonio, Texas 78229, USA.

\* Corresponding author: krsna@uthscsa.edu

**Abstract :**

*Fluorescence Lifetime Imaging Microscopy (FLIM) using multiphoton excitation techniques is now finding an important place in quantitative imaging of protein-protein interactions and intracellular physiology.  We review here the recent developments in multiphoton FLIM methods and also present a description of a novel multiphoton FLIM system using a streak camera that was developed in our laboratory.  We provide an example of a typical application of the system in which we measure the fluorescence resonance energy transfer between a donor/acceptor pair of fluorescent proteins within a cellular specimen.*

1.  **Introduction**

Detection of protein-protein interactions in living cells with high spatial and temporal resolution is a fundamental prerequisite for understanding cellular dynamics.[1] Visualization of specific probes by confocal and multiphoton imaging has improved spatial resolution in the context of cellular imaging in native, physiological conditions.[2,3] The implementation of Fluorescence Resonance Energy Transfer (FRET) detection in the optical microscope has extended the spatial detection limits of conventional fluorescence microscopy to a few nanometers – which is typically the magnitude of distance between the interacting proteins. Measurement of FRET requires that there exists a pair of fluorophores (donor and acceptor) between which there is non-radiative transfer of energy when they are in close proximity (~ 1-10 nm). The efficiency of energy transfer is determined by the distance separating the donor and acceptor molecules, the angular orientation of the fluorophores in space, overlap between the donor emission and acceptor excitation spectra and the quantum yield of the donor in the absence of acceptor.[4,5,6]

Many of the currently available FRET methods such as sensitized emission, acceptor photobleaching, and concentration dependent depolarization utilize intensity-based measurements. These require the use of optical filters for spectral separation of donor/acceptor excitation and emission and can suffer from problems of spectral crosstalk. In addition, it is difficult to regulate the relative concentration of the donor and acceptor fluorophores that in turn affects the efficiency of energy transfer thus further complicating the measurement of cell component interactions using intensity-based FRET approaches. These methods also yield time-averaged information from an ensemble of interacting proteins (or other molecules) rather than the actual temporal kinetics of these interactions in real-time. These constraints have led to the need to develop instrumentation that can measure FRET (protein-protein interactions) not only with high spatial resolution but also with a high degree of temporal resolution.

Fluorescence Lifetime Imaging Microscopy (FLIM) allows dynamic information from cells to be obtained with high temporal resolution and provides an extra dimension of information (the fluorescent lifetime) in cellular imaging.[7,8,9,10,11] FLIM methods are insensitive to the concentration of the fluorophore, yet can still provide distinct lifetimes characteristic of the individual fluorophores attached to each protein (e.g. donor and acceptor). By measuring the changes in lifetime of the donor - in the presence and in the absence of acceptor molecule, it is possible to estimate the energy transfer efficiency thereby allowing quantification of molecular interactions. By virtue of the fact that FLIM methods are devoid of aforementioned spectroscopic artifacts (as commonly encountered in intensity-based methods), the reliability in calculating quantitative FRET efficiencies is higher than in other methods. Besides FRET measurements, FLIM methods are also well suited to measure intracellular physiology such as pH, ion concentration etc., in real time.[12,13,14,15,16]

A variety of FLIM methods have been developed in the past decade for measuring various intracellular parameters by monitoring changes in fluorescence excited state lifetime. These have been broadly classified into two major methods: time domain and frequency domain FLIM. In frequency-domain methods, the specimen is excited by a sinusoidally modulated wave of a certain frequency (~ radio frequencies for nanosecond lifetimes; Fig.1a). The consequent fluorescence emission is also sinusoidally modulated but with different phase and amplitude because of the time lag induced by 'residence time' or the excited-state lifetime of the fluorophores. Generally a gain-modulated detector is used to determine the phase shift and amplitude demodulation from which the lifetimes of the fluorescent species are calculated.[17,18,19] Alternatively, in the case of time-domain methods, the specimen is illuminated with pulsed light and the subsequent fluorescence emission decay is recorded by fast detectors (Fig.1b). By fitting the observed emission decay to an appropriate model, the fluorescence lifetimes of the different fluorescent species within the specimen are extracted. Despite the availability of current FLIM methods for cell biological applications, a continual requirement that has pervaded this field is to achieve high spatial and temporal resolution in live cell applications. Conventional time-domain FLIM methods such as multigate detection, single photon counting etc., can measure lifetimes a short as a few hundred ps – a limitation imposed primarily by the detectors used in these systems. A recently developed imaging system determines the lifetime of fluorescence emission from multiphoton excitation imaging using a streak camera as the primary signal detector (Streak-FLIM). Although streak cameras have been used in studies of semiconductor phenomena and picosecond spectroscopy, they have not yet been employed for FLIM in biomedicalapplications.[20,21,22] In this context, the present multiphoton-Streak-FLIM system is unique and versatile, achieving excellent spatio-temporal resolution and, to the best of our knowledge, is the first of its kind developed for biomedical applications.

**2.     FLIM methodology**

The mono-exponential decay scheme of a fluorescent molecule from its excited state can be represented by:

$$I(t) = I_o \exp(-t/\tau) \qquad (1)$$

where $I(t)$ is the measured intensity and $I_o$ is the intensity at time t=0. The fluorescence lifetime, $\tau$, is characteristic of the molecule and does not depend on its concentration within the specimen or on the excitation light characteristics (intensity, path length in the specimen, etc.,). However, $\tau$ is sensitive to local environmental factors such as pH, ion concentration and is also sensitive to excited state processes such as quenching and energy transfer. As such, the measurement of changes in fluorescence lifetime under various experimental conditions functions as an indicator of alterations in intracellular physiology. For instance, FLIM offers a non-invasive way of measuring changes in $H^+$ or ion concentrations (e.g, $Ca^{2+}$ oscillations) as well as being a reliable way of probing protein-protein interactions with high temporal resolution. In this article, we briefly review the different time domain FLIM methods and describe in detail, the development and biomedical application of the StreakFLIM system.

*3.     Time Domain Detection Methods*

*3.1    Multi-gate detection*

In the simplest case of time-domain lifetime imaging, a gated micro-channel plate image intensifier is used in conjunction with a CCD imaging camera.[23,24] By gating the image intensifier at different intervals (with a specified time-window) along the exponential fluorescence decay profile, one can obtain a set of images of the actual decay of fluorescence during the excited state lifetime (Fig.1b). It has been found that the photon utilization and time resolution of the multi-gate detection approach are still limited by the detector performance. One can also not avoid background noise arising from scattered fluorescence emission when an imaging (area) detector is used in multi-gate detection. Regardless of these limitations, the multi-gate detection FLIM methodology has been found to be very successful with single photon excitation in the recent years.[25,26] For multiphoton applications, it is imperative to use point-scanning detectors to eliminate background noise.

### *3.2    Single photon counting*

Photon-counting detectors are well-suited for low-light level detection as well as for providing quantized pulses for every photon event.[27,28] This makes the measurement of lifetimes more accurate. However these detectors suffer from poor dynamic range as compared to their area detector counter parts. Regardless of this minor disadvantage, photon counting devices have been found to be very reliable in terms of photon economy, rapid lifetime determination and high temporal resolution. The limiting factor in achieving good temporal resolution is the transition-time spread (TTS) of fast photomultiplier tubes used in these systems.

### **3.3    Streak FLIM Imaging System**

The essential feature of the streak camera is that it converts an optical 2D image with spatial axes (x,y) into a streak image with temporal information and with the axes (x,t). A simplified picture of streak imaging principle is demonstrated in (Fig.2) and a brief description of the system is given below. For the complete technical details, see reference 30.

### 3.3.1 Light source

A mode-locked laser system (Coherent Inc. Model Mira 900: Titanium sapphire gain medium pumped by a 10W Verdi diode laser) was used in our experiments - providing ultrafast femtosecond pulses with a fundamental repetition rate 76 MHz. To obtain the variable repetition rates for both intensity based multiphoton imaging and for Streak-FLIM detection, a pulsepicker, synchronized with the Mira excitation source, was employed. Precise optical alignment of laser, the Pulsepicker and the FLIM optics was essential to ensure negligible absorption/distortion of the excitation laser beam. Optical alignment of the FLIM elements was done with respect to the optical axis of microscope objective.

### 3.3.2 *Microscope*

The imaging system base is an Olympus IX70 inverted microscope. For intensity based multiphoton images, the fs-pulsed laser light was scanned into the left camera port via a modified FluoView X scanhead. Non-descanned multiphoton emission was collected by a PMT in a custom designed housing mounted from the left side of the epi-fluorescence filter cassette. For Streak-FLIM imaging the fs-pulsed laser light was scanned into the right camera port by the Streak-FLIM unit. The fluorescence emission was directed back out through the same camera port and directed into the Streak camera (C-4334, Hamamatsu Photonics K.K.). A specially designed 63X (1.2 N.A, IR) water immersion objective was used for all the measurements reported in this paper. Experiments were done on either fixed cellular specimens mounted on standard microscope slides or on live cells grown on glass coverslips and maintained in appropriate media.

### *3.3.3 FLIM Detection System*

A microchannel plate (MCP) was used in the Streak-FLIM detection path. The Streakscope (Hamamatsu, C4334) has a temporal resolution ~50 ps and a very small photocathode dark current. This facilitates a high signal-to-noise ratio (SNR) for measuring even weak fluorescence signals. The electrical output of the MCP is converted to an optical output (streak image) on a phosphor screen. This image is then output to a fast CCD camera (Hamamatsu, ARGUS/HiSCA) that is fiber-optically coupled to the streakscope. An important feature of the HiSCA camera is that it employs a CCD chip that offers exceptionally high-speed and high sensitivity detection (~ 2% quantum efficiency). The 12-bit image depth enables even minute changes in fluorescence intensity to be detected. Variable sampling rate (1 to 24 frames/sec) and spatial resolution allow flexibility to detect fast photon events and low light signals. A maximum sampling rate of ~ 500 frames/sec can be achieved with the HiSCA camera for single wavelength fluorescence measurements which is more than ten times higher than the normal video-rate. Faster frame acquisition can be achieved by binning the pixels. Binning in the X-axis decreases spatial resolution while binning in the t (time) axis decreases the temporal resolution (Fig. 2B). As such a compromise between the speed of the data acquisition and required resolution must be achieved for each measurement. Unbinned, the maximum number of pixels of the CCD camera is 658 x 494, a maximum binning of 32 x 32 can be achieved. The FLIM data acquisition is governed via AquaCosmos software (Hamamatsu Photonics, K.K.) and the streakscope output signal is processed by AquaCosmos software to convert the exponential decay into a mean lifetime value on a per pixel basis thus creating a FLIM image display.

**3.4    System Calibration**

Standard solutions of various fluorescent dyes were used for calibration of the system and for optimizing relevant parameters for a typical measurement. Rhodamine 6G and Rose Bengal solutions were prepared in different solvents as detailed in Table 1. These dyes are reported to have mono-exponential fluorescent decays and therefore display a single lifetime.[29] The choice of standard dyes also allowed calibration of the system for measurement of lifetime values ranging from a few hundred picoseconds to a few nanoseconds. Table 1 gives the calculated lifetime values for these solutions obtained with the FLIM Strekscope system and compares these values with values reported in the literature for the same fluorophore solutions. There is a very good agreement between the calculated lifetimes with those in the literature, demonstrating the accuracy of lifetime determination of the Streak-FLIM system. Accuracy of the lifetime measurements is largely governed by the uniformity of the FLIM images. Lifetime images of the standard solutions that we employed exhibited a high degree of uniformity and thus provided high accuracy in the calculations of lifetime. Reproducibility between two measurements in the same specimen was found to be >99% in standard solutions.

However, for cellular specimens the variability in cellular expressions as well as inevitable photobleaching artifacts can restrict the reproducibility. A more detailed description of the instrument performance and signal/noise issues can be found in Reference 30.

*3.5    Biological applications*

Cyan – and Yellow fluorescent proteins (CFP/YFP) are mutated forms of the wild-type green fluorescent protein and are employed as a donor/acceptor pair in many FRET experiments.[31,32] Energy transfer can manifest in many ways such as reduction in donor emission fluorescence, sensitized fluorescence emission of the acceptor, reduction in donor lifetime etc.,  Of all the different FRET methods, measuring FRET by measuring the changes in donor lifetime has been reported to be more reliable owing to its relative insensitivity to spectroscopic artifacts commonly encountered in intensity based measurements. In this section, we give a typical cellular application of the Streak FLIM system in studying caspase activity in single cells. Baby Hamster Kidney (BHK) cells were transiently transfected with a mitochondrially targeted DNA construct (mC2Y) that links CFP and YFP by an amino acid encoding caspase-2 (VDVAD) recognition sequence.[33] Under identical conditions of growth and transfection, a batch of cells was exposed to the broad-range caspase activator tert-butyl hydrogen peroxide (tBOOH: 50 μM) for 4 hours at 37$^o$C in order to induce oxidative stress in mitochondria. Experiments were carried out under identical imaging conditions for all of the specimens. It is expected that the close proximity of CFP and YFP molecules in the intact caspase-2 substrate will lead to significant energy transfer in the mC2Y specimens. However, induction of caspase-2 activity in tBOOH-treated mC2Y specimens will lead to cleavage of the substrate, thereby altering the spatial relationship between the CFP and YFP molecules and decreasing the extent of energy transfer in these specimens. We measured lifetimes of the CFP in both the tBOOH-treated and untreated cells and calculated the change in energy transfer efficiency in these two situations.

For measurements of CFP fluorescence lifetime the fs-pulsed laser was tuned to 840 nm. To select for emission from CFP only the fluorescent signal passed through a 480/30 emission filter (Chroma Inc.) placed before the streakscope. Donor lifetime images were calculated using AquaCosmos software. Figure 3 shows the (CFP) lifetime images of the mC2Y and mC2Y + tBOOH specimens. The mC2Y specimen shows punctate mitochondrial expression of CFP whereas there is an observable morphological change in the tBOOH treated specimen. This morphological change (the cell becomes more rounded), together with the diffuse fluorescence in Figure 3b, can be attributed to oxidative stress induced by tBOOH and the consequent induction of apoptosis in the cell. As can be seen from the lifetime histogram, the mean lifetime of the mC2Y specimen is around 2 ns whereas that of the tBOOH-treated mC2Y specimen is around 2.28 ns. This increase in lifetime is a clear manifestation of loss of energy transfer between CFP and YFP molecules in the tBOOH-treated mC2Yspecimen as compared to the untreated-mC2Y specimen where the molecules are held intact by the intervening caspase-2 peptide sequence. This indicates that there is a significant mitochondrial caspase activity induced by oxidative stress. As a positive control for FRET measurements, we used DNA constructs encoding mitochondrially targeted CFP-polyglycine-YFP (mCGY) in BHK fibroblast cells and imaged under conditions identical to that for our experimental studies (mC2Y).  Both these specimens exhibited average lifetime of ~2.0 ns.  Upon treatment with caspase activator, tBOOH, only mC2Y specimen showed a 15% increase in average lifetime whereas the mCGY specimen showed less than 5% change in lifetime.  This proved that what we see as change in lifetime in mC2Y specimen upon tBOOH treatment is essentially due to caspase activity.  It is also intriguing to note that the lifetime distribution in the tBOOH-treated specimen is much broader than that of the untreated mC2Y specimen.  This can be interpreted as representing non-uniform caspase cleavage activity within the cell.  Some of the caspase-2 substrates have been cleaved at the time when the lifetime measurement was made while other substrates still remain intact.  This would cause a wide distribution in the extent of energy transfer spatially within the cell and manifest itself as a wider distribution in the measured lifetime in the tBOOH-treated specimen.  Although it would be interesting to analyze these histograms by multicomponent lifetime models, only single component gaussian model was adopted in the present analysis.  Furthermore, at 4 hours post-tBOOH treatment, only about 50% of the cells showed changes in lifetime upon tBOOH treatment.  Different cells in an ensemble would be at different stages of apoptosis at any one time.  Since the fixation of the cells was done at a single time point after the tBOOH treatment, it is expected that not all the cells would exhibit apoptotic changes in lifetime measurement.  It is also possible that there exists a substantial population of uncleaved caspase-2 substrate even in the tBOOH-treated specimen as this substrate is being constitutively expressed.

   The energy transfer efficiency can be calculated from the expression :
$E = 1 - (\tau_{DA}/\tau_D)$ where $\tau_{DA}$ and $\tau_D$ are the lifetimes of the donor in the presence and in the absence of acceptor respectively.  In the present case, $\tau_{DA}$ and $\tau_D$ represent the mean lifetimes of CFP in the untreated and tBOOH-treated specimens respectively. From the fitting of mean lifetime histograms in both the specimens, the change in energy transfer efficiency in the above example was calculated to be ~15%.

We have taken care to ensure that the calculated lifetime values are not affected by photobleaching artifacts or autofluorescence contributions. Imaging conditions were optimized for minimal photobleaching so as to get reproducible lifetime values from the same region of the cell in consecutive scans. It has been observed earlier that autofluorescence from fibroblast cells have typical lifetime values around 2.2 ns.[15] We measured the streak images of non-transfected cells and found that the maximum fluorescence intensity contribution from autofluorescence is less than 5% of that observed in transfected cells. We have thresholded the StreakFLIM image intensity values above this value in our lifetime calculations in order to remove the ambiguity of a possible contribution of autofluorescence to the observed lifetime. It is also worth mentioning that autofluorescence contributions have been found to be less using multiphoton excitation.[34]

The above demonstration of reliable FRET measurement using multiphoton lifetime imaging combined with a streak camera opens up new possibilities for quantitative imaging of protein-protein interactions in living cellular systems. Since there is only a requirement to measure donor lifetime in all the specimens, the problem of spectral crosstalk is overcome. However, it is advisable to measure the lifetime of donor in the absence of acceptor under identical imaging conditions that can then serve as a control for non-FRET conditions.

In recent years there has been an upsurge in the quantitative imaging of molecular interactions in addition to the intensity-dependent visualization of intracellular probes. Improvements in fluorescence techniques such as confocal, multiphoton microscopy and other energy transfer methods have yielded valuable spatial/spectral information pertinent to submicron molecular interactions. Lifetime imaging methods offer improved temporal resolution in fluorescence imaging thereby incorporating an additional experimental parameter. Although single photon FLIM techniques have been in use for many years, it is expected that the combination of multiphoton excitation and FLIM methods will yield an enhancement in spatio-temporal resolution and reduce overall phototoxicity and thereby improve our understanding of molecular interactions.

**Table 1 Legend**

The standard solutions were prepared by dissolving the dye powders in different solvents and diluted stocks were prepared for calibration. Calculated lifetime values did not change with the concentration of the dye in the nominal range of a few hundred nM to a few hundred μM. All the measurements were carried out at around 200 μM concentration for system calibration. Raw streak images were obtained and the mean lifetimes were calculated by the methods described in the text. Monoexponential decays were assumed for all the calibration samples as reported in literature and the values reported here are all mean lifetime values. Standard deviations were obtained by fitting the lifetime histograms to normal distribution.

**Table 1**:   Measured Lifetimes in the Streak-FLIM system

| Samples | | Measured Lifetimes | | Literature | |
|---|---|---|---|---|---|
| | | $\tau$ (ns) | Standard deviation (ns) | $\tau$ (ns) | Reference |
| Calibration Probes | Rhodamine 6G / Ethanol | 2.97 | 0.14 | 3.0 | 29 |
| | RoseBengal / Acetone | 2.22 | 0.13 | 2.4 | 29 |
| | RoseBengal / Ethanol | 0.74 | 0.04 | 0.8 | 29 |

**Figure Legends**

**Figure 1 :** *(A) Schematic of the principle of frequency domain lifetime imaging. $\omega$ is the modulation frequency. For single component lifetime of fluorescent species, the lifetimes $\tau(\phi)$ and $\tau(M)$ are the same but differ if there are more than one lifetime component. (B) Schematic of the principle of the time-domain lifetime imaging. Gated image intensifiers are set at different time windows ( w1,w2) with a defined time interval ($\Delta t$) and the fluorescent lifetime is calculated according to the expression given above.*

**Figure 2:** *(A) Streak FLIM imaging system is composed of the laser system ( Coherent Mira 900 and Coherent Pulsepicker 9200), Optics ( Olympus IX-70 inverted microscope, FLIM optics) and the Streak system (trigger unit, Streakscope, HiSCA Argus camera). (B) A schematic of the principle of streak imaging is shown. A single line in the optical image with intensity information at every point (a) is converted by the streakscope into a streak image (b) with spatial information as the horizontal axis and time as the vertical axis. Every point (containing fluorescent molecules) along this line has an exponential decay profile as shown in (c). By scanning along the entire optical image ((x,y)plane), similar exponential decay profiles are produced corresponding to every single point in the entire plane. These decay profiles are stored and then numerically processed to give a lifetime image as shown in (d). See text for more technical details.*

**Figure 3:** *Fluorescence Lifetime images of baby hamster kidney cells transfected with mitochondrially targeted DNA construct (mC2Y) that links CFP and YFP via an amino acid sequence encoding caspase-2 recognition site. (a) untreated mC2Y (b) mC2Y treated with 50 µM tBOOH for 4 hours (c) lifetime histogram for both the specimens. Laser output from Mira 900 (76 MHz rep rate; 840 nm wavelength; ~200 fs pulsewidth) is rendered to the pulsepicker which steps down the repetition rate to 500 kHz optimized for the streakscope triggering. The average power at the entrance of the FLIM optics is 8 mW. The 12-bit streak images were obtained in the 1x32 binning condition. Exposure times for imaging (~195 ms per single line) were optimized to minimize photobleaching during data acquisition. The lifetime images were calculated from the raw streak images by AquaCosmos software (Hamamatsu). An intensity threshold criterion of about 10% of the maximum dynamic range was used in calculation of lifetime images and the final images were median filtered. Lifetime histograms shown were plotted for the entire cells. Mean lifetime of the mC2Y specimen was 1.98 ns whereas that of tBOOH-treated mC2Y specimen was 2.28 ns.*

**References**

[1] Methods in Cellular Imaging, Ed. A. Periasamy, (Oxford University Press, NewYork) 2001

[2] J. B. Pawley (1995) *Handbook of biological Confocal Microscopy* Plenum Press, New York.

[3] Konig.K,(2000) Multiphoton microscopy in life sciences, *J.Microsc.* **200**, 83-104

[4] Clegg, R.M (1995) Fluorescence resonance energy transfer, *Curr Opin Biotechnol*, **6**, 103 -110

[5] Selvin, P.R (2000) The renaissance of fluorescence resonance energy transfer *Nature Struct.Biol.* **7**, 730 – 734.

[6] Krishnan, R.V, Varma,R and Mayor.S (2001), Fluorescence methods to probe nanometer scale organization of molecules in living cell membrane. *J.Fluorescence,* **11**, 211 – 226.

[7] Lakowicz, J. R. (1996). "Emerging applications of fluorescence spectroscopy to cellular imaging: lifetime imaging, metal-ligand probes, multi-photon excitation and light quenching." Scanning Microsc Suppl **10**: 213-24.

[8] Oida, T., Y. Sako, et al. (1993). "Fluorescence lifetime imaging microscopy (flimscopy). Methodology development and application to studies of endosome fusion in single cells." Biophys J **64**(3): 676-85.

[9] Squire, A. and P. I. Bastiaens (1999). "Three dimensional image restoration in fluorescence lifetime imaging microscopy." J Microsc **193 ( Pt 1)**: 36-49.

[10] Squire, A., P. J. Verveer, et al. (2000). "Multiple frequency fluorescence lifetime imaging microscopy." J Microsc **197 ( Pt 2)**: 136-49

[11] Clayton, A. H., Q. S. Hanley, et al. (2002). "Dynamic fluorescence anisotropy imaging microscopy in the frequency domain (rFLIM)." Biophys J **83**(3): 1631-49.

[12] Szmacinski, H. and J. R. Lakowicz (1995). "Possibility of simultaneously measuring low and high calcium concentrations using Fura-2 and lifetime-based sensing." Cell Calcium **18**(1): 64-75.

[13] Gadella, T. W., Jr. and T. M. Jovin (1995). "Oligomerization of epidermal growth factor receptors on A431 cells studied by time-resolved fluorescence imaging microscopy. A stereochemical model for tyrosine kinase receptor activation." J Cell Biol **129**(6): 1543-58.

[14] Verveer, P. J., F. S. Wouters, et al. (2000). "Quantitative imaging of lateral ErbB1 receptor signal propagation in the plasma membrane." Science **290**(5496): 1567-70

[15] van Zandvoort, M. A., C. J. de Grauw, et al. (2002). "Discrimination of DNA and RNA in cells by a vital fluorescent probe: lifetime imaging of SYTO13 in healthy and apoptotic cells." Cytometry **47**(4): 226-35

[16] Elangovan, M., R. N. Day, et al. (2002). "Nanosecond fluorescence resonance energy transfer-fluorescence lifetime imaging microscopy to localize the protein interactions in a single living cell." J Microsc **205**(Pt 1): 3-14

[17] French, T., P. T. So, et al. (1997). "Two-photon fluorescence lifetime imaging microscopy of macrophage-mediated antigen processing." J Microsc **185 (Pt 3)**: 339-53.

[18] Verveer, P. J., A. Squire, et al. (2001). "Improved spatial discrimination of protein reaction states in cells by global analysis and deconvolution of fluorescence lifetime imaging microscopy data." J Microsc **202**(Pt 3): 451-6

[19] Verveer, P. J., A. Squire, et al. (2000). "Global analysis of fluorescence lifetime imaging microscopy data." Biophys J **78**(4): 2127-37

[20] T. Tahara and H.O. Hamaguchi, Appl. Spectrosc. **47**, 391 ( 1993).

[21] M. Heya, S. Fujioka, H. Shiraga, N. Miyanaga and T. Yamanaka, Rev. Sci. Instrum. **72**, 755 ( 2001).

[22] U. Neuberth, L. Walter, G. von Freymann, B. Dal Don, H. Kalt, M. Wegener, G. Khitrova and H.M. Gibbs, Appl. Phys. Lett. **80**, 3340 ( 2002).

[23] M. Straub and S.W. Hell, Appl.Phys.Lett. **73**, 1769 (1998)

[24] Gerritsen, H.C, M.A.H. Asselbergs, A.V. Agronskaia and W.G.J.A.H.M. Van Sark (2002) Fluorescence lifetime imaging in scanning microscopes: acquisition speed, photon economy and lifetime resolution, J.Microsc. **206**, 218-224

[25] Cole, M. J., J. Siegel, et al. (2001). "Time-domain whole-field fluorescence lifetime imaging with optical sectioning." J Microsc **203**(Pt 3): 246-57

[26] Sytsma, J, J.M.Vroom, C.J.de Grauw, H.C.Gerritsen,(1998) Time-gated lifetime

imaging and micro-volume spectroscopy using two-photon excitation, J.Microsc. **191,** 39-51

[27] Schonle A, M.Glatz, and S.W.Hell, (2000) Four-dimensional multiphoton microscopy with time-correlated single-photon counting, Appl.Opt. **39** : 6308-6311

[28] Becker W, A.Bergmann and G.Weiss (2002) Lifetime imaging with the Zeiss LSM-510, Proc SPIE **4620**, 30-35

[29] A. Periasamy, P. Wodnicki, X.F.Wang, S. Kwon, G.W. Gordon and B. Herman, Time-resolved fluorescence lifetime imaging microscopy using a picosecondpulsed tunable dye laser system. Rev. Sci. Instrum. **67**, 3722 (1996).

[30] Krishnan R.V, H.Saitoh, H.Terada, V.E. Centonze and B.Herman, Development of multiphoton fluorescence lifetime imaging microscopy (FLIM) system using a streak Camera, Rev.Sci.Instrum.(In press)

[31] Tavare J.M, L.M Fletcher and G.I.Welsh (2001) Using green fluorescent protein to study intracellular signaling , J.Endocrinol. **170**, 297-306 and references therein;

[32] Harpur, A. G., F. S. Wouters, et al. (2001). "Imaging FRET between spectrally similar GFP molecules in single cells." Nat Biotechnol **19**(2): 167-9.

[33] Mahajan.N.P, D.C.Harrison-Shostak, J.Michaux and B.Herman (1999), Novel mutant green fluorescent protein protease substrates reveal the activation of specific caspases during apoptosis, Chemistry and Biology, **6**,: 401-409

[34] Konig, K., P. T. So, et al. (1996). "Two-photon excited lifetime imaging of autofluorescence in cells during UVA and NIR photostress." J Microsc **183 ( Pt 3)**: 197-204.

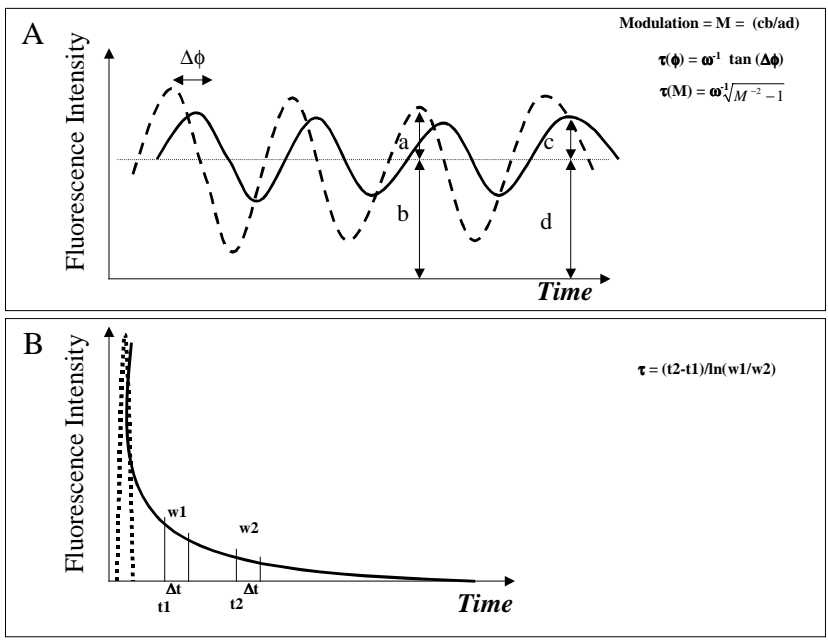

Fig.1

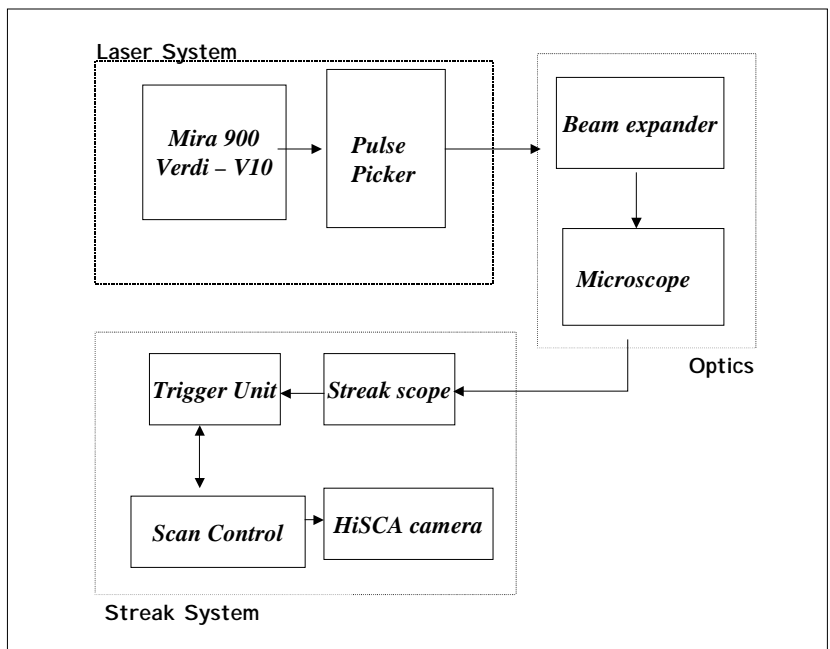

Fig.2A

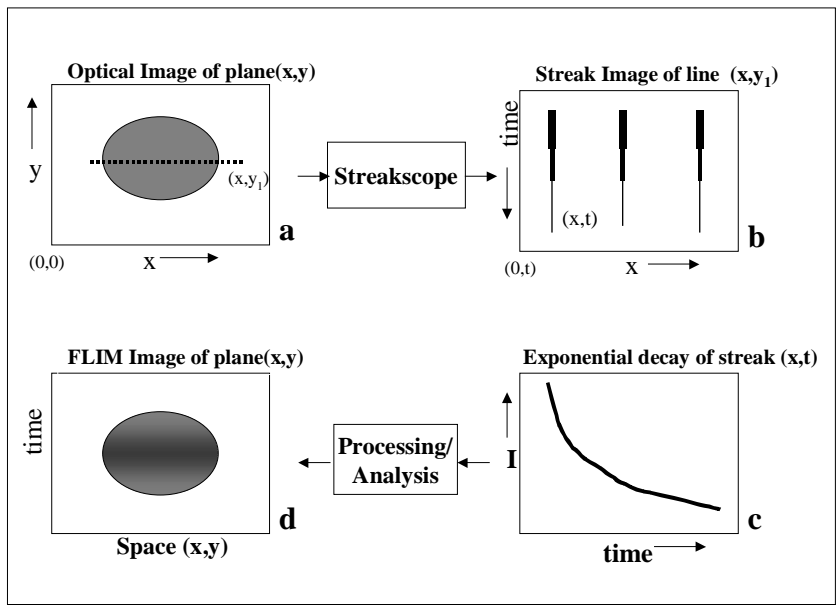

Fig.2B

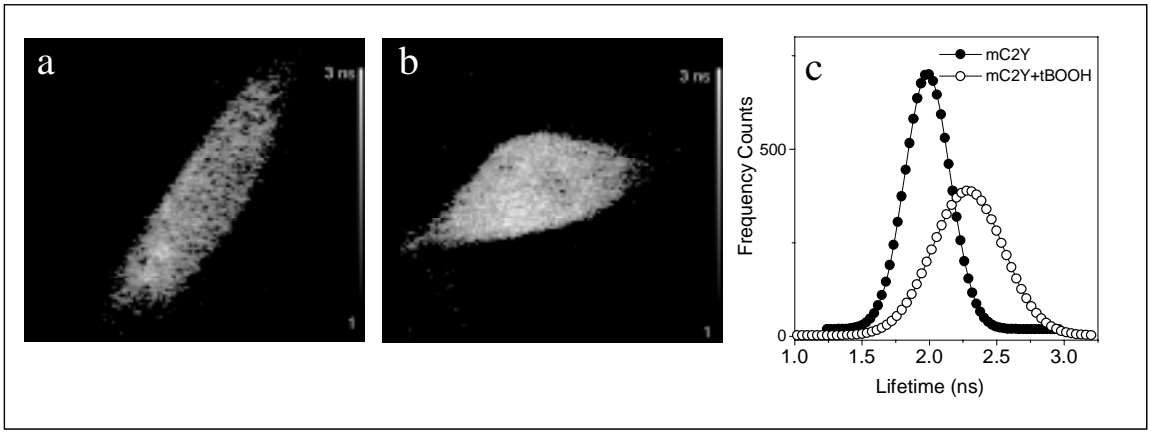

Fig.3